\documentclass[runningheads]{llncs}
\usepackage[hidelinks]{hyperref}
\usepackage{graphicx}%
\usepackage[T1]{fontenc}
\usepackage{multirow}%
\usepackage{amsmath,amssymb,amsfonts}%
\usepackage{mathrsfs}%
\usepackage[title]{appendix}%
\usepackage{xcolor}%
\usepackage{textcomp}%
\usepackage{manyfoot}%
\usepackage{booktabs}%
\usepackage{algorithm}%
\usepackage{algorithmicx}%
\usepackage{algpseudocode}%
\usepackage{listings}%
\usepackage{orcidlink}
\usepackage{enumitem}
\setlist{nosep}

\begin{document}

\title{Edit3DGS: Unified Framework for Dynamic Head Editing via 2D Instruction-Guided Diffusion and 3D Gaussian Splatting}
\titlerunning{Edit3DGS: Unified Framework for Dynamic Head Editing}

\author{Duy-Dat Tran \inst{1,2} \orcidlink{0009-0004-3136-2589} \and Trung-Nghia Le \inst{1,2}\orcidlink{0000-0002-7363-2610}\thanks{Corresponding author.}}

\authorrunning{Duy-Dat Tran and Trung-Nghia Le}

% Affiliations
\institute{University of Science, VNU-HCM, Ho Chi Minh, Vietnam \and Vietnam National University, Ho Chi Minh, Vietnam \\
\email{22C11058@student.hcmus.edu.vn}, \email{ltnghia@fit.hcmus.edu.vn}}

\maketitle

\begin{abstract}
We present Edit3DGS, a unified framework for dynamic 3D head editing that integrates 2D instruction-guided diffusion with 3D Gaussian splatting. Unlike prior approaches that separately address frame-based edits or static 3D reconstruction, our method couples semantic controllability in the image domain with photorealistic, temporally consistent 3D representations. Given an input video, editable facial regions are masked and modified using a text-conditioned diffusion model to support fine-grained operations such as expression transformation, attribute modification, and appearance refinement. The edited frames are then aggregated through 3D Gaussian splatting to produce a coherent, high-fidelity avatar that preserves both identity and motion dynamics. To enforce consistency, Edit3DGS incorporates multi-view batch editing and lightweight inpainting strategies that recover lost expressions across timesteps. Experimental results demonstrate that our framework enables controllable, artifact-free head editing with smooth temporal transitions, offering practical applications in virtual avatars, immersive communication, film production, and interactive media.

\end{abstract}

\section{Introduction}\label{intro}

3D reconstruction of the human head has become a central topic in computer vision and graphics, enabling the creation of high-quality digital replicas with diverse geometric and appearance characteristics. Recent advances have extended this capability beyond static geometry, supporting photorealistic rendering, dynamic capture, and smooth animation. Among these developments, 3D Gaussian Splatting (3DGS) \cite{3dgs} has emerged as a powerful representation, leveraging anisotropic Gaussian primitives to model complex surfaces and volumetric effects with both high fidelity and real-time performance. By combining the efficiency of point-based rendering with the expressiveness of continuous radiance fields, 3DGS surpasses Neural Radiance Fields (NeRF) \cite{nerf} in rendering quality and speed for novel view synthesis. As a result, 3DGS has been widely adopted in human head reconstruction and animation (head avatars) \cite{3dgsavatars,qian2024gaussianavatars,chen2024monogaussianavatar,saito2024relightable,xiang2024flashavatar}.

Despite rapid progress, editing 3DGS-based avatars, such as modifying expressions, attributes, or styles, remains limited. At the same time, diffusion-based text-to-image models \cite{stablediffusion,podell2023sdxl} have transformed 2D semantic editing, enabling intuitive manipulation of visual content with natural language prompts and fine-grained control. This paradigm has motivated text-driven 3D editing, where the goal is to extend the same flexibility and expressiveness to volumetric representations. However, directly integrating 2D diffusion models into 3DGS pipelines presents major challenges. Maintaining spatial and temporal consistency across views and frames is particularly difficult when dealing with dynamic facial features and subtle expressions. Naïve editing approaches often yield inconsistencies, visual artifacts, or identity drift.

To address these challenges, we propose Edit3DGS, a unified framework that combines semantic guidance from 2D instruction-driven diffusion with the structural fidelity of 3D Gaussian splatting. By introducing batch multi-view editing and a lightweight inpainting strategy, Edit3DGS enforces spatial coherence across views and temporal smoothness across frames, enabling controllable and photorealistic edits of dynamic head avatars. To the best of our knowledge, this is the first method that achieves editable 3DGS-based head avatars while preserving structural integrity and rendering quality.

Extensive experiments on the NeRSemble dataset \cite{kirschstein2023nersemble} validate the effectiveness of our approach. Qualitative results show that Edit3DGS produces smooth, photorealistic edits across challenging expressions and viewpoints, while quantitative evaluations using CLIP-based metrics confirm that our method achieves competitive or superior alignment with text prompts compared to GaussianAvatar-Editor \cite{liu2025gaussianavatar}, a recent state-of-the-art method. Together, these results highlight the practicality of Edit3DGS for real-world applications in avatar creation, film production, and interactive media.

Our main contributions are summarized as follows:

\begin{itemize}
\item We introduce Edit3DGS, a simple yet efficient unified framework for editing dynamic 3DGS-based head avatars by combining instruction-guided 2D diffusion with Gaussian fitting.
\item We propose multi-view batch editing and an auto-generated inpainting mask to preserve expression and maintain temporal consistency across frames.
\item We demonstrate that Edit3DGS achieves high-quality, temporally consistent, and controllable edits across tasks such as novel view synthesis, self-reenactment, and cross-reenactment.
\end{itemize}

\section{Related Work} \label{related}
\subsection{3D Editing} 

Recent advancements in deep learning have catalyzed a paradigm shift in 3D content creation. Traditional workflows relied heavily on labor-intensive manual modeling, but the field has moved toward intuitive, data-driven approaches that leverage learning-based methods. These methods operate on both explicit and implicit representations. For explicit forms, approaches such as MeshCNN \cite{hanocka2019meshcnn} employ graph neural networks to enable mesh-based manipulation and style transfer. In parallel, implicit representations have gained traction, most notably Neural Radiance Fields (NeRFs) \cite{nerf}, which encode geometry and appearance within a continuous neural function.

Building on this foundation, text-driven editing has emerged as a powerful paradigm. Methods such as Instruct-NeRF2NeRF \cite{haque2023instruct} and TIP-Editor \cite{zhuang2024tip} allow users to perform large-scale modifications or fine-grained changes using natural language prompts. This shift toward generative, text-conditioned editing democratizes 3D content creation by abstracting technical complexity and supporting more intuitive, creative workflows. DreamFusion \cite{poole2022dreamfusion} advanced this direction by introducing Score Distillation Sampling (SDS), which leverages a pre-trained text-to-image diffusion model as a prior to optimize a NeRF representation into a 3D asset from scratch. Subsequent research has extended these principles to editing existing 3D scenes. DreamEditor \cite{zhuang2023dreameditor}, for example, proposed a framework which represents scenes as mesh-based neural fields, allowing targeted manipulation of specific regions while preserving the integrity of unedited areas. This decoupled design significantly enhances both the quality and controllability of 3D scene manipulation.

\subsection{Gaussian Head Avatar Reconstruction}
The remarkable rendering speed and visual fidelity of 3D Gaussian Splatting (3DGS) \cite{3dgs} has led to a surge of research applying it to photorealistic human head modeling and animation. These approaches primarily vary in how they use a 3D Morphable Model (3DMM) like FLAME \cite{flame} to rig and deform the Gaussians. Several methods, including GaussianAvatars \cite{qian2024gaussianavatars} and FlashAvatar \cite{xiang2024flashavatar}, utilize the 3DMM mesh structure to drive Gaussian movement. GaussianAvatars directly rigs Gaussians to the mesh triangles, achieving high fidelity but lacking control over non-facial features like hair and teeth, as these are not explicitly modeled by FLAME. FlashAvatar addresses the uneven distribution issue by sampling Gaussians in UV space, maintaining a uniform density and achieving faster training. However, it requires a separate MLP to model offsets for details beyond the surface, and its initialization is highly sensitive.

Other works focus on integrating Gaussian deformation with advanced neural techniques. Saito et al. \cite{saito2024relightable} also samples in UV space, but uses a Conditional VAE to learn latent expression distributions. This allows them to achieve photorealistic relighting for hair and skin, though it involves a more complex appearance processing pipeline. Similarly, MonoGaussianAvatar \cite{chen2024monogaussianavatar} introduces a dedicated Gaussian deformation field to capture novel shape structures, like accessories, achieving flexible topology. A key limitation across these methods is their reliance on 3DMM priors, which can cause a failure to capture extreme or unusual expressions that fall outside the model's domain.

\section{Proposed Method}\label{method}
\begin{figure}[t!]
\includegraphics[width=\textwidth]{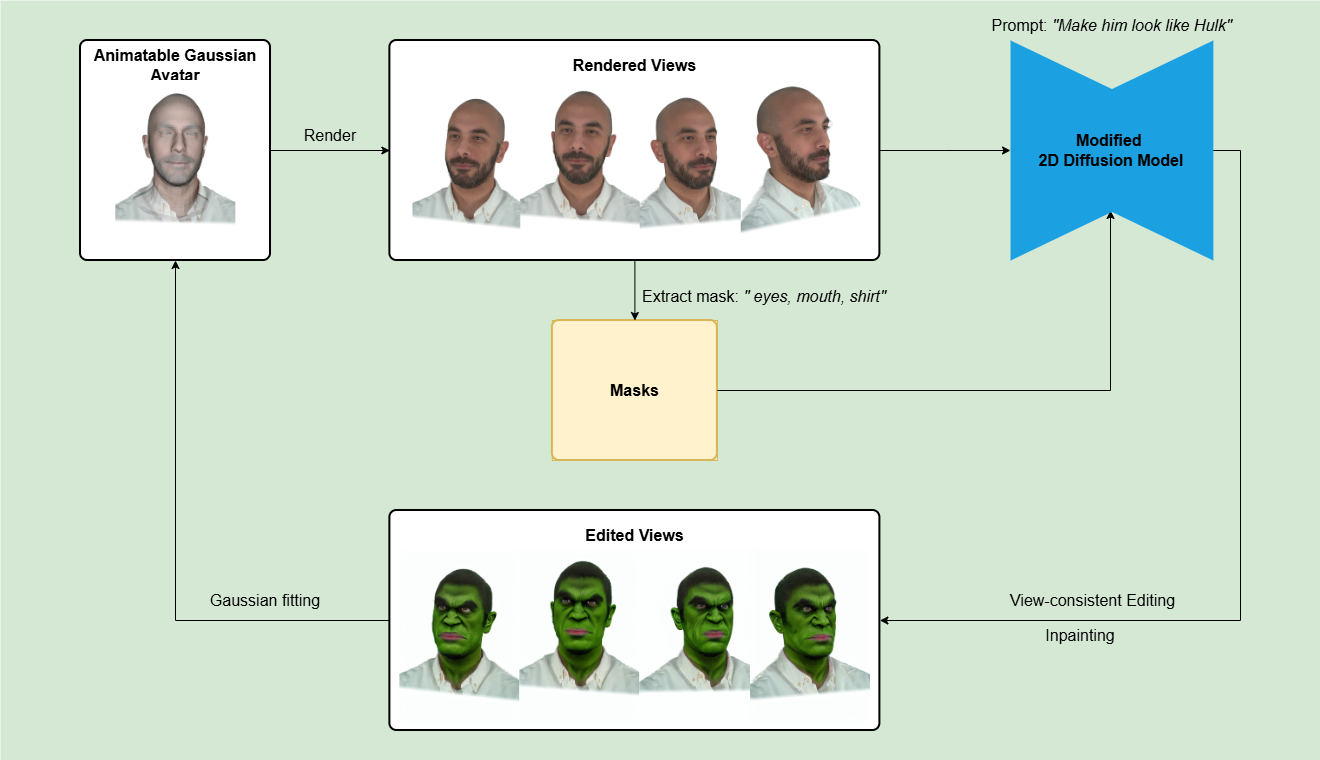}
\caption{Overview of the proposed Edit3DGS framework. The core components are multi-view batch editing and mask-based refinement, which together ensure both spatial consistency across views and preservation of facial expressions. At each timestep, selected camera views are rendered and edited using instruction-guided diffusion. The edited results across multiple timesteps are then aggregated to update the original Gaussian model through 3D fitting. After several iterations, the refined 3D avatar can be reprocessed through 2D editing to propagate changes to previously unseen timesteps.} \label{figframework}
\end{figure}

The overview of our proposed method is shown in Fig.~\ref{figframework}. Starting from a Gaussian head avatar generated by GaussianAvatars \cite{qian2024gaussianavatars}, we select a set of images rendered from chosen cameras and timesteps. These images are then processed by a text-guided diffusion model, producing an edited dataset that serves as input for Gaussian fitting. Through this process, the head model is iteratively updated to align with the newly edited dataset.

\subsection{GaussianAvatars}
The main idea of GaussianAvatars \cite{qian2024gaussianavatars} is to fit a set of 3D Gaussian splats to a parametric morphable face model, specifically the FLAME \cite{flame} mesh, enabling precise control over expressions, pose, and viewpoint. Each Gaussian is associated with its parent triangle on the mesh, and the system jointly optimizes both the morphable model and the Gaussian parameters in an end-to-end manner. This design combines the photorealistic rendering strengths of 3D Gaussian Splatting with the expressiveness and controllability of a rigged mesh, leading to superior quality and animation control compared to existing methods. For this reason, we adopt GaussianAvatars as the base model for 3D head reconstruction within our Edit3DGS framework.

\subsection{Edit Through Gaussian Fitting}\label{subsec2}

If we can obtain a set of multi-view consistent edited images from renders of a 3D Gaussian model, then the model itself can be directly updated through Gaussian fitting. This makes the training process both fast and efficient due to the nature of 3DGS-based techniques. The key challenge, however, is how to generate such consistent edited views. A straightforward solution is to apply text-driven editing models, such as Instruct-Pix2Pix \cite{brooks2023instructpix2pix}, to each view independently. While simple, this approach produces results that often vary across timesteps and camera angles. As a consequence, the resulting 3D model reconstructed from these edits suffers from inconsistency, blur, and artifacts. Therefore, the central problem is to enhance 2D text-guided editing methods so that each view accounts for both spatial and temporal constraints.

\subsection{Spatially Consistent Editing}\label{subsubsec2}

Given a set of views $V$ rendered from a 3D Gaussian model at timestep $t$, our goal is to ensure that the edited views remain consistent with one another, as in the original dataset. This requires the editing model to capture the mutual dependencies between views. To achieve this, we adopt the idea of the multi-view editing process proposed in DGE \cite{chen2024dge}. In our framework, $V$ is treated as a batch input to the 2D diffusion model, and editing proceeds in two steps: key-view editing and feature injection.

\textbf{Key-view editing.} During each denoising step, a subset of views from $V$ is randomly selected as key views. These views are edited simultaneously with spatial-temporal attention. Specifically, we extend a Stable Diffusion-based model \cite{stablediffusion} with spatial-temporal attention blocks, enabling each view to incorporate contextual information from the others. This step extracts key information that guides subsequent editing.

\textbf{Feature injection.} Once key views are edited, their features are propagated to all other views to ensure batch-wide consistency. Correspondences between key views and non-key views are established using epipolar constraints. By comparing matching visual features extracted from different layers of the denoising network, we transfer key features across views. This process enforces coherent edits across the batch. Details of this mechanism can be found in the original DGE paper \cite{chen2024dge}.

\subsection{Temporally Consistent Editing}

The render-edit-aggregate pipeline introduced by Instruct-NeRF2NeRF \cite{haque2023instruct} has demonstrated effectiveness in reducing inconsistency for static 3D scene editing, and it has also been applied in 3DGS-based settings \cite{chen2024gaussianeditor}. By iteratively editing a rendered view and propagating changes to subsequent frames, temporal consistency gradually emerges after several iterations. Although Instruct-NeRF2NeRF was designed for static scenes, we adapt this principle to dynamic head avatars.

We observe that, at the same viewpoint, the primary differences between two renders at different timesteps lie in expressions and poses of the head model. Current text-guided diffusion editors such as Instruct-Pix2Pix \cite{brooks2023instructpix2pix} often produce nearly identical faces across frames, failing to preserve complex expressions. As a result, they frequently lose critical emotional details, particularly around the eyes and mouth. To address this limitation, we propose a simple yet effective solution: using inpainting with latent diffusion models to recover and preserve facial expressions.

Our approach works as follows. During editing, the input images are mapped back into latent space at different noise levels. Masks are then applied to isolate latent vectors corresponding to the eyes and mouth of the original faces. These latent vectors are re-injected into the denoising process at later steps, ensuring that the original expressions reappear in the final edits. Beyond expression preservation, this inpainting strategy also supports flexible local editing of specific facial regions.

\section{Experiments}\label{experiments}
\subsection{Setup}
\subsubsection{Implementation Details.} We used a 3D head avatar trained by GaussianAvatars \cite{qian2024gaussianavatars} pipeline using a set of multi-view dataset as our base mode for editing. As for the text-based image editor, we used Instruct-Pix2Pix \cite{brooks2023instructpix2pix} for easily comparison with other current methods. For inpainting process, we used the combining of SAM 2 \cite{ravi2024sam} and Grounding DINO \cite{liu2024grounding} to automatically extract the masks which contain only eyes and mouth of the rendered images. The model was trained using Adam optimizer with a learning rate of $1e - 2$, running for $10 \times n$ iteration per editing, with $n$ being the number of timesteps/batches in the considered dataset. In the inference phase, we can directly achieve the renders with novel poses, expressions and viewpoint of the head model under the guidance of FLAME \cite{flame} parameters. 

\subsubsection{Dataset.} The experiments utilized the multi-view video dataset NeRSemble \cite{kirschstein2023nersemble}, which consists of 11 video sequences for each subject, and each video sequence contains approximately 150 frames of different expressions and movements. The dataset are downscale to $802 \times 550$, this results in not only less time spent in creating new dataset but also ensuring the high fidelity of the edited head avatar. We also left one video sequence and one camera view for quantitative evaluation as in the original GaussianAvatars. 

\subsubsection{Evaluation.} We provide both qualitative and quantitative evaluations of our
method. For qualitative evaluation, we follow the original paper to evaluate the edited animatable Gaussian models in three main use cases: novel view rendering, self-reenactment and cross-identity reeenactment. As for quantitative evaluation, we calculate the common CLIP Text-Image Direction Similarity (CLIP-S) \cite{brooks2023instructpix2pix,gal2022stylegan} and CLIP Direction Consistency (CLIP-C) \cite{haque2023instruct} under all said circumstances. 

\subsubsection{Baseline.} To the best of our knowledge, only one prior work (e.g., GaussianAvatars-Editor \cite{liu2025gaussianavatar}) addresses the problem of editable 3DGS-based dynamic head models. This method also adopts the same baseline as ours and achieves strong results. Therefore, we use GaussianAvatars-Editor \cite{liu2025gaussianavatar} as the primary baseline for comparison.

\subsection{Qualitative Results}

\begin{figure}[t!]
\centering
\includegraphics[width=\textwidth]{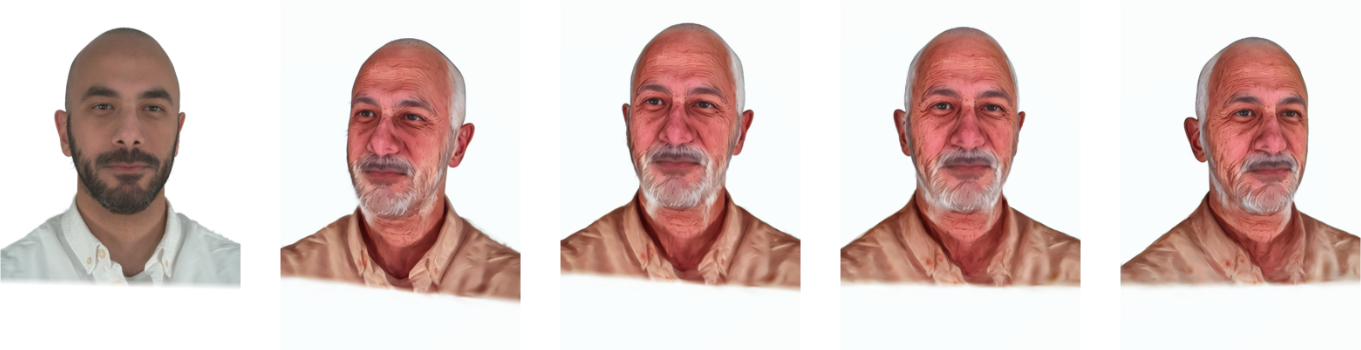}
\caption{Qualitative results of the novel view synthesis experiment. 
Given the text prompt \textit{``Make him look older''}, our method generates consistent and high-quality edits across different viewpoints, while preserving identity and structural details.} 
\label{novelviewfig}
\end{figure}

\subsubsection{Novel View Rendering}

As illustrated in Fig.~\ref{novelviewfig}, when provided with a 3D Gaussian avatar and the prompt \textit{``Make him look older''}, our method is able to produce photorealistic edits that remain highly consistent across multiple viewpoints. Edit3DGS preserves facial structure and ensures smooth transitions across camera angles. The generated results clearly demonstrate that the integration of 2D diffusion guidance with Gaussian fitting yields not only semantic accuracy (e.g., age-related changes in skin texture and facial attributes) but also spatial coherence across views.

\subsubsection{Self and Cross-Identity Reenactment}

To further evaluate the versatility of our method, we conduct both self-reenactment and cross-identity reenactment experiments.  

\begin{figure}[t!]
\centering
\includegraphics[width=\textwidth]{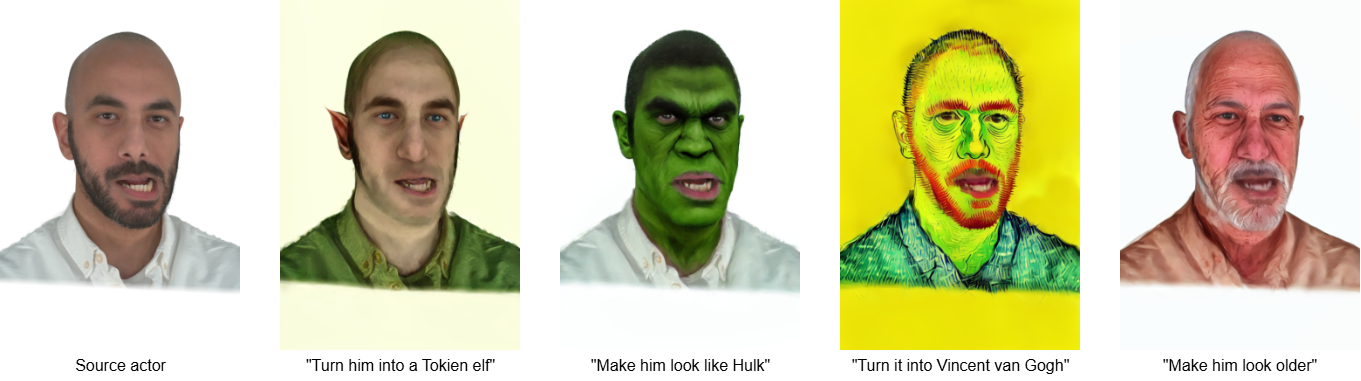}
\caption{Qualitative results of the self-reenactment experiment. 
The edited avatar is driven by unseen expressions and poses from the same actor. 
The text prompts corresponding to each edit are shown below the images.} 
\label{selfreenactres}
\end{figure}

In the self-reenactment setting (Fig.~\ref{selfreenactres}), the edited avatar is animated using unseen expressions and poses from the same actor. Our method produces smooth, temporally coherent sequences that retain fine-grained details such as subtle wrinkles, eye gaze, and lip motion. Importantly, the edits remain stable across diverse expressions, confirming the robustness of Edit3DGS under challenging motion dynamics.  

\begin{figure}[t!]
\centering
\includegraphics[width=\textwidth]{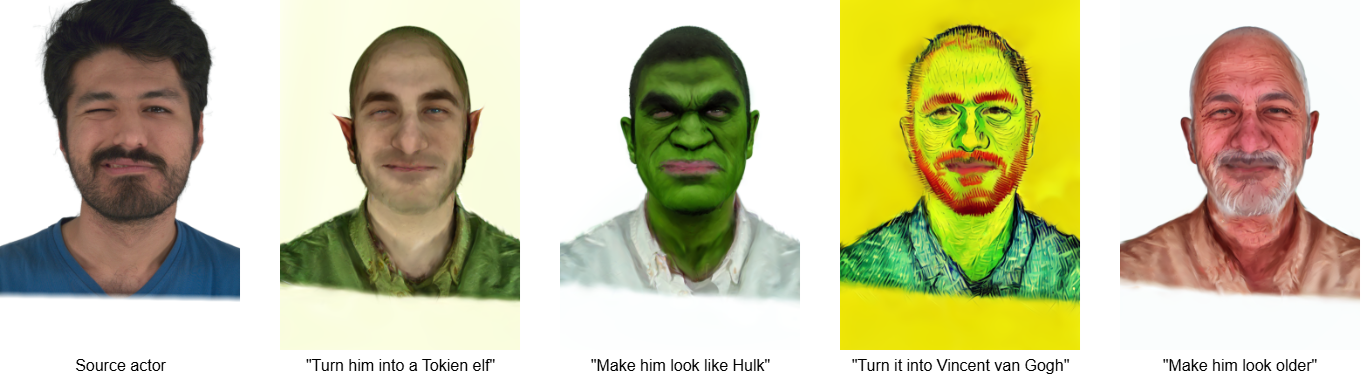}
\caption{Qualitative results of the cross-identity reenactment experiment. 
Expressions and poses from a different actor are transferred to animate the edited avatar. 
Text prompts corresponding to each edit are provided below the images.} 
\label{crossreenactres}
\end{figure}

In the cross-identity reenactment setting (Fig.~\ref{crossreenactres}), expressions and poses from a different actor are transferred to drive the edited avatar. This is a more difficult scenario since identity preservation and motion fidelity must be balanced simultaneously. Our method successfully preserves the identity of the edited subject while accurately transferring complex expressions such as smiles, raised eyebrows, or mouth openings from the source actor.

\subsection{Quantitative results}

The quantitative comparison in Table~\ref{tab:quantitative_comparison} reports results across three tasks using the CLIP-S \cite{brooks2023instructpix2pix,gal2022stylegan} and CLIP-C \cite{haque2023instruct} metrics. Overall, the results highlight that the performance gap between our proposed \textbf{Edit3DGS} and GaussianAvatar-Editor \cite{liu2025gaussianavatar} is very narrow, with both methods achieving highly competitive outcomes.

Across all six metrics, the absolute differences remain small. Our method achieves slightly higher CLIP-S scores for novel view rendering and cross-identity reenactment, suggesting stronger semantic alignment between the edited images and the guiding text prompts. Conversely, GaussianAvatar-Editor shows marginally better CLIP-C scores in some categories, indicating slightly more directional consistency in its edits.

These numerical variations, however, are modest and unlikely to translate into noticeable differences in perceptual quality. Both methods deliver strong performance across all tasks, confirming their effectiveness for dynamic head editing. Importantly, when considered alongside our qualitative evaluations, Edit3DGS demonstrates an advantageous balance: while maintaining metrics on par with the state of the art, it produces visually more coherent and expressive results, particularly in challenging scenarios involving complex expressions and large pose variations. This suggests that Edit3DGS not only matches existing approaches quantitatively but also offers practical improvements in edit controllability and temporal stability.

\begin{table}[t]
\centering
\caption{Quantitative comparison with previous work. We report CLIP-based metrics for three tasks: novel view rendering, self-reenactment, and cross-identity reenactment. }
\label{tab:quantitative_comparison}
\setlength{\tabcolsep}{3pt}
\begin{tabular}{lcccccc}
\toprule
& \multicolumn{2}{c}{\textbf{Novel view}} & \multicolumn{2}{c}{\textbf{Self-reenact.}} & \multicolumn{2}{c}{\textbf{Cross-reenact.}} \\
\cmidrule(lr){2-3} \cmidrule(lr){4-5} \cmidrule(lr){6-7}
& \textbf{CLIP-S}$\uparrow$ & \textbf{CLIP-C}$\uparrow$ 
& \textbf{CLIP-S}$\uparrow$ & \textbf{CLIP-C}$\uparrow$ 
& \textbf{CLIP-S}$\uparrow$ & \textbf{CLIP-C}$\uparrow$ \\
\midrule
GA-Editor~\cite{liu2025gaussianavatar} & 0.258 & \textbf{0.978} & \textbf{0.076} & 0.954 & 0.072 & \textbf{0.965} \\
Edit3DGS (Ours) & \textbf{0.269} & 0.969 & 0.071 & \textbf{0.957} & \textbf{0.074} & 0.945 \\
\bottomrule
\end{tabular}
\end{table}

\subsection{Ablation study}
\subsubsection{Inpainting.} 
\begin{figure}[t!]
\centering
\includegraphics[width=0.8\textwidth]{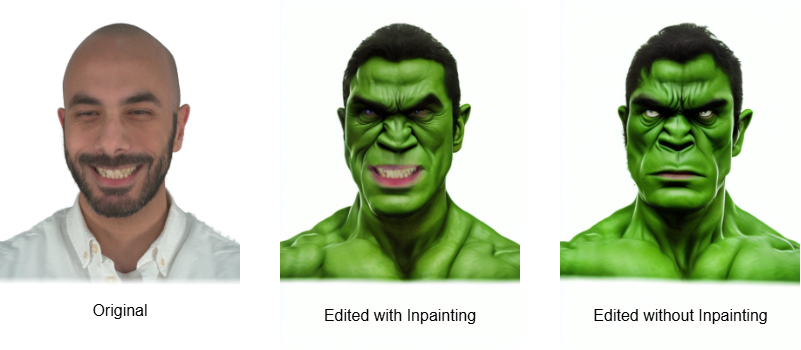}
\caption{Comparison between edited results with and without inpainting process.} 
\label{abl1}
\end{figure}

We conduct an experiment to validate the importance of our proposed inpainting method by disabling it in the pipeline. The edited results are shown in Fig \ref{abl1}. The results display that without the inpainting process, the edited views can not retain the original expression of the head, particularly the details around eyes and mouth, which will worsen the performance in animation. 

\section{Conclusion}\label{conclusion}

In this paper, we introduced a pioneering framework for editable 3DGS-based head avatars. Our approach, {Edit3DGS}, enables dynamic 3D head editing by combining multi-view batch editing with instruction-guided diffusion and a lightweight masked inpainting strategy. This design allows fine-grained, temporally consistent edits while preserving photorealism and structural integrity.

Despite its effectiveness, the method still inherits certain limitations from the FLAME model, which cannot perfectly capture fine details such as teeth and hair. In addition, performance is influenced by the choice of 2D diffusion editors, leading to variability in speed and stability. Addressing these challenges opens promising avenues for future research, such as integrating stronger 3D priors, developing more robust expression modeling, and designing domain-adaptive 2D–3D editing pipelines.

Beyond technical contributions, Edit3DGS has the potential to impact a wide range of real-world applications. These include creating controllable avatars for immersive communication and social interaction, producing high-quality assets for film and gaming, enabling personalized virtual assistants in AR/VR, and advancing interactive media where user-driven customization is key. By bridging semantic controllability in 2D with high-fidelity dynamic 3D representations, Edit3DGS lays a solid foundation for the next generation of expressive and editable digital humans.

\begin{credits}

% This research is supported by research funding from Faculty of Information Technology, University of Science, Vietnam National University - Ho Chi Minh City.
\subsubsection{\ackname}
This research is funded by Vietnam National Foundation for Science and Technology Development (NAFOSTED) under Grant Number 102.05-2023.31. This research used the GPUs provided by the Intelligent Systems Lab at the Faculty of Information Technology, University of Science, VNU-HCM.
\end{credits}
\bibliographystyle{splncs04}
\bibliography{sample-base}

@String{Computer = "{IEEE} Computer" }

@String{Springer = "Springer-Verlag" }

@article{3dgs,
  title={3D Gaussian splatting for real-time radiance field rendering.},
  author={Kerbl, Bernhard and Kopanas, Georgios and Leimk{\"u}hler, Thomas and Drettakis, George},
  journal={ACM Trans. Graph.},
  volume={42},
  number={4},
  pages={139--1},
  year={2023}
}

@article{nerf,
  title={Nerf: Representing scenes as neural radiance fields for view synthesis},
  author={Mildenhall, Ben and Srinivasan, Pratul P and Tancik, Matthew and Barron, Jonathan T and Ramamoorthi, Ravi and Ng, Ren},
  journal={Communications of the ACM},
  volume={65},
  number={1},
  pages={99--106},
  year={2021},
  publisher={ACM New York, NY, USA}
}

@inproceedings{3dgsavatars,
  title={3dgs-avatar: Animatable avatars via deformable 3d gaussian splatting},
  author={Qian, Zhiyin and Wang, Shaofei and Mihajlovic, Marko and Geiger, Andreas and Tang, Siyu},
  booktitle={CVPR},
  pages={5020--5030},
  year={2024}
}

@inproceedings{qian2024gaussianavatars,
  title={Gaussianavatars: Photorealistic head avatars with rigged 3d gaussians},
  author={Qian, Shenhan and Kirschstein, Tobias and Schoneveld, Liam and Davoli, Davide and Giebenhain, Simon and Nie{\ss}ner, Matthias},
  booktitle={CVPR},
  pages={20299--20309},
  year={2024}
}

@inproceedings{xiang2024flashavatar,
  title={Flashavatar: High-fidelity head avatar with efficient gaussian embedding},
  author={Xiang, Jun and Gao, Xuan and Guo, Yudong and Zhang, Juyong},
  booktitle={CVPR},
  pages={1802--1812},
  year={2024}
}

@inproceedings{saito2024relightable,
  title={Relightable gaussian codec avatars},
  author={Saito, Shunsuke and Schwartz, Gabriel and Simon, Tomas and Li, Junxuan and Nam, Giljoo},
  booktitle={CVPR},
  pages={130--141},
  year={2024}
}

@inproceedings{chen2024monogaussianavatar,
  title={Monogaussianavatar: Monocular gaussian point-based head avatar},
  author={Chen, Yufan and Wang, Lizhen and Li, Qijing and Xiao, Hongjiang and Zhang, Shengping and Yao, Hongxun and Liu, Yebin},
  booktitle={ACM SIGGRAPH 2024 Conference Papers},
  pages={1--9},
  year={2024}
}

@article{stablediffusion,
  author       = {Robin Rombach and
                  Andreas Blattmann and
                  Dominik Lorenz and
                  Patrick Esser and
                  Bj{\"{o}}rn Ommer},
  title        = {High-Resolution Image Synthesis with Latent Diffusion Models},
  journal      = {CoRR},
  volume       = {abs/2112.10752},
  year         = {2021},
  url          = {https://arxiv.org/abs/2112.10752},
  eprinttype    = {arXiv},
  eprint       = {2112.10752},
  bibsource    = {dblp computer science bibliography, https://dblp.org}
}

@article{podell2023sdxl,
  title={Sdxl: Improving latent diffusion models for high-resolution image synthesis},
  author={Podell, Dustin and English, Zion and Lacey, Kyle and Blattmann, Andreas and Dockhorn, Tim and M{\"u}ller, Jonas and Penna, Joe and Rombach, Robin},
  journal={arXiv preprint arXiv:2307.01952},
  year={2023}
}

@article{flame,
  title={Learning a model of facial shape and expression from 4D scans},
  author={Tianye Li and Timo Bolkart and Michael J. Black and Hao Li and Javier Romero},
  journal={ACM Transactions on Graphics (TOG)},
  year={2017},
  volume={36},
  pages={1 - 17},
  url={https://api.semanticscholar.org/CorpusID:9882090}
}

@article{hanocka2019meshcnn,
  title={Meshcnn: a network with an edge},
  author={Hanocka, Rana and Hertz, Amir and Fish, Noa and Giryes, Raja and Fleishman, Shachar and Cohen-Or, Daniel},
  journal={ACM Transactions on Graphics (ToG)},
  volume={38},
  number={4},
  pages={1--12},
  year={2019},
  publisher={ACM New York, NY, USA}
}

@inproceedings{haque2023instruct,
  title={Instruct-nerf2nerf: Editing 3d scenes with instructions},
  author={Haque, Ayaan and Tancik, Matthew and Efros, Alexei A and Holynski, Aleksander and Kanazawa, Angjoo},
  booktitle={Proceedings of the IEEE/CVF international conference on computer vision},
  pages={19740--19750},
  year={2023}
}

@article{zhuang2024tip,
  title={Tip-editor: An accurate 3d editor following both text-prompts and image-prompts},
  author={Zhuang, Jingyu and Kang, Di and Cao, Yan-Pei and Li, Guanbin and Lin, Liang and Shan, Ying},
  journal={ACM Transactions on Graphics (TOG)},
  volume={43},
  number={4},
  pages={1--12},
  year={2024},
  publisher={ACM New York, NY, USA}
}

@article{poole2022dreamfusion,
  title={Dreamfusion: Text-to-3d using 2d diffusion},
  author={Poole, Ben and Jain, Ajay and Barron, Jonathan T and Mildenhall, Ben},
  journal={arXiv preprint arXiv:2209.14988},
  year={2022}
}

@inproceedings{zhuang2023dreameditor,
  title={Dreameditor: Text-driven 3d scene editing with neural fields},
  author={Zhuang, Jingyu and Wang, Chen and Lin, Liang and Liu, Lingjie and Li, Guanbin},
  booktitle={SIGGRAPH Asia 2023 Conference Papers},
  pages={1--10},
  year={2023}
}

@article{liu2025gaussianavatar,
  title={GaussianAvatar-Editor: Photorealistic Animatable Gaussian Head Avatar Editor},
  author={Liu, Xiangyue and Luo, Kunming and Li, Heng and Zhang, Qi and Liu, Yuan and Yi, Li and Tan, Ping},
  journal={arXiv preprint arXiv:2501.09978},
  year={2025}
}

@inproceedings{brooks2023instructpix2pix,
  title={Instructpix2pix: Learning to follow image editing instructions},
  author={Brooks, Tim and Holynski, Aleksander and Efros, Alexei A},
  booktitle={CVPR},
  pages={18392--18402},
  year={2023}
}

@inproceedings{chen2024dge,
  title={Dge: Direct gaussian 3d editing by consistent multi-view editing},
  author={Chen, Minghao and Laina, Iro and Vedaldi, Andrea},
  booktitle={European Conference on Computer Vision},
  pages={74--92},
  year={2024},
  organization={Springer}
}

@inproceedings{chen2024gaussianeditor,
  title={Gaussianeditor: Swift and controllable 3d editing with gaussian splatting},
  author={Chen, Yiwen and Chen, Zilong and Zhang, Chi and Wang, Feng and Yang, Xiaofeng and Wang, Yikai and Cai, Zhongang and Yang, Lei and Liu, Huaping and Lin, Guosheng},
  booktitle={CVPR},
  pages={21476--21485},
  year={2024}
}

@article{kirschstein2023nersemble,
  title={Nersemble: Multi-view radiance field reconstruction of human heads},
  author={Kirschstein, Tobias and Qian, Shenhan and Giebenhain, Simon and Walter, Tim and Nie{\ss}ner, Matthias},
  journal={ACM Transactions on Graphics (TOG)},
  volume={42},
  number={4},
  pages={1--14},
  year={2023},
  publisher={ACM New York, NY, USA}
}

@article{ravi2024sam,
  title={Sam 2: Segment anything in images and videos},
  author={Ravi, Nikhila and Gabeur, Valentin and Hu, Yuan-Ting and Hu, Ronghang and Ryali, Chaitanya and Ma, Tengyu and Khedr, Haitham and R{\"a}dle, Roman and Rolland, Chloe and Gustafson, Laura and others},
  journal={arXiv preprint arXiv:2408.00714},
  year={2024}
}

@inproceedings{liu2024grounding,
  title={Grounding dino: Marrying dino with grounded pre-training for open-set object detection},
  author={Liu, Shilong and Zeng, Zhaoyang and Ren, Tianhe and Li, Feng and Zhang, Hao and Yang, Jie and Jiang, Qing and Li, Chunyuan and Yang, Jianwei and Su, Hang and others},
  booktitle={European conference on computer vision},
  pages={38--55},
  year={2024},
  organization={Springer}
}

@article{gal2022stylegan,
  title={Stylegan-nada: Clip-guided domain adaptation of image generators},
  author={Gal, Rinon and Patashnik, Or and Maron, Haggai and Bermano, Amit H and Chechik, Gal and Cohen-Or, Daniel},
  journal={ACM Transactions on Graphics (TOG)},
  volume={41},
  number={4},
  pages={1--13},
  year={2022},
  publisher={ACM New York, NY, USA}
}
\end{document}